\documentclass{aastex61}

\received{\today}
\revised{}
\accepted{}
\submitjournal{ApJ}

\shorttitle{Variability of solar wind power spectrum}
\shortauthors{Roberts et al.}
\usepackage{graphicx}
\begin{document}

\title{Variability of the magnetic field power spectrum in the solar wind at electron scales}

\correspondingauthor{Owen Wyn Roberts}
\email{Owen.Roberts@esa.int}

\author{Owen Wyn Roberts}
\affil{European Space Agency, ESA-ESTEC SCI-S, Keplerlaan 1, 2201 AZ Noordwijk, The Netherlands }
\author{O. Alexandrova}
\affil{LESIA, Observatoire de Paris, PSL Research University, CNRS, UPMC Universite Paris 06, Universite Paris-Diderot, 5 Place Jules Janssen, F-92190 Meudon, France}
\author{P. Kajdi\v{c}}
\affil{Institutio de Geof\'isica, Universidad Nacional Aut\'onoma de M\'exico, Mexico City, M\'exico}
\author{L. Turc}
\affil{European Space Agency, ESA-ESTEC SCI-S, Keplerlaan 1, 2201 AZ Noordwijk, The Netherlands }
\affil{Department of Physics, University of Helsinki, Helsinki, Finland}
\author{D. Perrone}
\affil{European Space Agency, ESA-ESAC, Madrid, Spain}
\author{C.P. Escoubet}
\affil{European Space Agency, ESA-ESTEC SCI-S, Keplerlaan 1, 2201 AZ Noordwijk, The Netherlands }
\author{A. Walsh}
\affil{European Space Agency, ESA-ESAC, Madrid, Spain}



\begin{abstract}
At the electron scales the power spectrum of solar-wind magnetic fluctuations can be highly variable and the dissipation mechanisms of the magnetic energy into the various particle species is under debate. In this paper we investigate data from the Cluster mission's STAFF Search Coil magnetometer when the level of turbulence is sufficiently high that the morphology of the power spectrum at electron scales can be investigated. The Cluster spacecraft sample a disturbed interval of plasma where two streams of solar wind interact. Meanwhile, several discontinuities (coherent structures) are seen in the large scale magnetic field, while at small scales several intermittent bursts of wave activity (whistler waves) are present. Several different morphologies of the power spectrum can be identified: (1) two power laws separated by a break (2) an exponential cutoff near the Taylor shifted electron scales and (3) strong spectral knees at the Taylor shifted electron scales. These different morphologies are investigated by using wavelet coherence, showing that in this interval a clear break and strong spectral knees are features which are associated with sporadic quasi parallel propagating whistler waves, even for short times. On the other hand, when no signatures of whistler waves at $\sim 0.1-0.2f_{ce}$ are present, a clear break is difficult to find and the spectrum is often more characteristic of a power law with an exponential cutoff.

\end{abstract}




\section{Introduction} \label{Intro}
Turbulence is characterized by disordered fluctuations in parameters over a large range of scales. In contrast to turbulence in a classical fluid, plasma turbulence contains several different important characteristic scales, which affect the shape of the power spectra \citep[see the reviews of][]{Bruno2013,Alexandrova2013,Kiyani2015}. To investigate plasma turbulence it is advantageous to investigate a plasma where these characteristic scales are much larger than our instruments. The solar wind offers such an environment \citep{Tu1995,Bruno2013}.

The typical solar wind magnetic field spectrum contains several different scales where different physics dominate. At large scales a fluid description is appropriate and the power spectrum of magnetic fluctuations shows frequency dependence of $f^{-1}$ and consists of uncorrelated Alfv\'en waves \citep{Belcher1971}. This is followed by a break near the correlation length \citep{Matthaeus2005}, and a steepening to a Kolmogorov like inertial range with a scaling of $f^{-5/3}$ \citep{Tu1995} where magnetohydrodynamics is valid, these scales are often termed the fluid scales. However as we descend to proton characteristic scales, (e.g. gyroradius/cyclotron frequency/ inertial length) the fluid description of a plasma is no longer valid and kinetic effects become important. This is often termed the ion kinetic range and a spectral break is seen near either the shifted proton Larmor radius ($f_{\rho_{p}}=V_{sw}/2\pi \rho_{p}$) or inertial length ($f_{d_{i}}=V_{sw}/2\pi d_{i}$) \citep{Bourouaine2012,Alexandrova2013,Chen2014a}. The spectral break is then followed by a steepening \citep{Leamon1998,Smith2006,Alexandrova2008c}, where a transition range between the fluid and sub-ion kinetic scales is often observed \citep{Howes2008b,Sahraoui2010a,Alexandrova2013}. 

The nature of this region (and regions at smaller scales) has been the subject of many recent spirited debates.  The critical balance hypothesis \citep{Goldreich1995} supposes that the characteristic timescales of the turbulence i.e. the Alfv\'en timescale (linear) and the eddy turnover time (nonlinear) continuously evolve towards being equal. Therefore in a nonlinear turbulent cascade the linear terms are of the same order as the nonlinear terms and the system may retain some of the properties of linear physics \citep{Klein2012,Howes2014,Howes2015}. Consequently, several have argued that the nonlinear cascade could have some properties of a superposition of linear wave solutions of the Vlasov equation \cite{Klein2012,TenBarge2012}. Although it should be noted that a superposition of plane waves cannot reproduce intermittency \citep{Howes2014} which is frequently observed in the solar wind (e.g \cite{Bruno2013}). However, it is possible  that the nonlinear fluctuations can keep the polarization properties of linear waves, (e.g. Lacombe et al. 2017 Accepted in APJ.)  

Anti-correlations of magnetic field strength and density are routinely observed at fluid \citep{Howes2012} and ion kinetic scales \citep{Yao2011} which are characteristic of kinetic slow waves (KSW) \citep{Zhao2014,Narita2015} or pressure balanced structures. In intervals of fast solar wind ion cyclotron waves may also be present at scales larger than the spectral break \citep{He2011,Podesta2011,Smith2012,Roberts2015} and can have an influence on the shape of the spectrum \citep{Lion2016}. To continue down past the spectral break to sub ion scales there are two candidates for waves that could be used to describe the cascade: kinetic Alfv\'en waves (or KAW) \citep{Bale2005,Sahraoui2010a,Salem2012,Chen2013,Podesta2013,Kiyani2013}, or magnetosonic/Bernstein waves \citep{Stawicki2001,Li2001a,Perschke2013,Perschke2014}.  

A different interpretation is that these scales are populated by a series of nonlinear fluctuations \citep{Perschke2014,Narita2017} or coherent structures which can also have an effect on the turbulent power spectrum \citep{Lion2016}. These may include discontinuities, current sheets or magnetic vortices \citep{Osman2011,Roberts2016,Perrone2016}, which can exist at fluid scales such as the MHD Alfv\'en vortex \citep{Lion2016,Roberts2016,Perrone2016} and to sub ion scales such as the current sheet observed by \cite{Perri2012}. Turbulent fluctuations could also represent a complex mixture of fluctuations which are wave-like and those which are characteristic of coherent structures \citep{Roberts2013,Karimabadi2013,Roberts2015a,Perschke2016,Lion2016,Roberts2017}.

At even smaller scales, close to characteristic scales of electrons (hereafter electron scales) the nature of the spectrum is more difficult to discern. Several observations of the spectrum have revealed a scaling close to -2.8 \citep{Alexandrova2009,Sahraoui2010a,Alexandrova2012,Sahraoui2013}. Beyond this range the morphology of the spectrum is unclear, the timescales and amplitudes of the fluctuations at these scales are small requiring an instrument with both a high sampling rate and a high signal to noise ratio (SNR). Several studies have been performed in the solar wind using the {\it{Cluster}} mission's \citep{Escoubet1997,Escoubet2001} spatio-temporal analysis of field fluctuations (STAFF) instrument \citep{Cornilleau-Wehrlin2003} which consists of both a spectral analyzer (SA) and a search coil magnetometer (SCM). When the {\it{Cluster}} spacecraft are in burst mode the SCM magnetic field data are sampled at 450Hz, and fluctuations often have sufficient amplitudes for a suitable SNR.

Due to both the aforementioned instrumental limitations and the lack (until recently with the Magnetospheric Multiscale Mission) of multi-spacecraft data with very small separations, the nature of the fluctuations at electron scales is somewhat unclear. One possibility is that a kinetic Alfv\'en wave cascade persists down to these scales \citep{Howes2011,Chen2013}. A recent statistical study of $\delta \mathbf{B}$ anisotropy by \emph{Lacombe et al. (2017, in press APJ)} shows that the KAW is the dominant mode at sub ion scales. However, at electron scales the plasma becomes more compressible, and in some conditions the kinetic slow wave may be important to account for the compressibility. Alternatively the KAW itself may become more compressible near the electron inertial length as suggested by \cite{Passot2017,Chen2017}, which could also account for the increase in compressibility. 

It should be noted that the linear damping rate of KAWs is significant \citep{Narita2015} and is sensitive to the plasma parameters, (e.g. plasma $\beta$ or ion to electron temperature ratio \citep{Howes2006}). It is possible that a KAW could be damped by electrons before reaching electron scales.  Moreover damping of these waves may be compounded by occurring at each stage in the cascade \citep{Podesta2010}. Alternatively dissipation could occur in localized coherent structures such as current sheets; in the study of \cite{Perri2012} current sheets were found with {\it{Cluster}} data when two of the four spacecraft were only 20 km apart in a projected direction. 

%A multi-spacecraft wave analysis method with the newly operational Magnetospheric Multiscale Mission (MMS) consisting of four spacecraft with separations of 15 km suggested that perpendicular magnetosonic/whistler waves are present at these scales \citep{Narita2016b}.    

Two main models have been proposed to model the `typical' magnetic field power spectral density (PSD) at these scales. The first model proposes that the PSD can be modeled as a power law with an exponential cutoff (Eq. \ref{exp} \cite{Alexandrova2009,Alexandrova2012}). This is similar to that in a neutral fluid (e.g. \cite{Chen1993}) and contains three free parameters. A statistical study of 100 individual ten minute spectra allowing the coverage from both STAFF-SA and STAFF-SCM found that all spectra could be fitted with the exponential model. The alternative to this model is the break model which consists of two separate power law spectra separated by a spectral break (Eq \ref{break}). \cite{Alexandrova2012} found that this model could only fit 30 of the 100 spectra with comparable errors to the exponential model, with those giving a mean scaling of -2.8 and -3.9.    

\begin{equation}
PSD(f)=Af^{-  \alpha} \exp{\left(-  f/ f_{c} \right)}
\label{exp}
\end{equation}

\begin{equation}
PSD(f)=A_{1}f^{- \alpha_{1}}(1 -  H(f -  f_{b}))+A_{2}f^{-  \alpha_{2}}H(f -  f_{b})
\label{break}
\end{equation}

In a separate statistical study of the solar wind spectrum \cite{Sahraoui2013} found better agreement with the break model (Eq \ref{break}) which consists of five free parameters. This study examined 620 spectra lasting $\sim10s$ and a mean scaling of -4.0 was found. \cite{Sahraoui2013} also argued that the exponential cutoff observed by \cite{Alexandrova2009,Alexandrova2012} was due to the averaging effects since they were computed over a time of 10 minutes, where plasma parameters (and the corresponding characteristic scales) can vary within the time interval. However this was refuted for the observations of \cite{Alexandrova2012}; the solar wind speed and the electron Larmor radius were not found to vary significantly during the times the spectra were taken. Another possibility to overcome the issues with low SNR is to investigate turbulence in the Earth's magnetosheath; here the fluctuation amplitudes of the turbulence are larger yielding greater SNR (e.g. \cite{Alexandrova2008a}).

Several efforts to simulate the PSD at these scales have been made. However, similar requirements of high time resolution and SNR make such simulations challenging. Simulations based on gyrokinetics \citep{Howes2011} have succeeded in reproducing similar PSDs for both electric and magnetic fields to observations in an interval of slow solar wind (e.g. \cite{Bale2005}). These simulations neglect effects of cyclotron resonance, which may be an important channel for the dissipation of magnetic energy especially in the fast solar wind (e.g. \cite{Bruno2014,Telloni2015,Roberts2015a,Lion2016}). Another approach is the Particle in Cell (PIC) approach, however at small scales numerical shot noise becomes an issue making the SNR low unless a sufficiently large number of particles are provided in the simulation (e.g. \cite{Camporeale2011}). Several particle in cell simulations of electron scale turbulence have been performed yielding similar spectral slopes to those observed in-situ \citep{Camporeale2011,Chang2011,Wan2012,Haynes2014,Wan2015}. Moreover, the PIC simulations of \cite{Roytershteyn2015} recover an exponential spectral shape. 

Some spectra at electron scales have been observed with atypical features such as large spectral knees near the shifted electron characteristic scales \citep{Sahraoui2013a,Lacombe2014}. These enhancements have been shown by \cite{Lacombe2014} to correspond to coherent right hand polarized fluctuations which propagate along the mean magnetic field direction, consistent with Magnetosonic/whistler waves. Although these events are relatively rare they contain a significant proportion of the power at certain frequencies when they do occur. A separate analysis using electric field data from the ARTEMIS spacecraft also found waves with dispersion relations consistent with parallel whistler waves \citep{Stansby2016}. Such fluctuations were also found to be present in the magnetosheath \citep{Matteini2017}.

The goal of this paper will be to investigate the shape of the power spectrum at electron scales for a single event where a large spectral knee is present in the Fourier spectra calculated between 2009-01-31 04:52-04:53UT on {\it{Cluster 2}}, presented in Fig 1d of \cite{Sahraoui2013a}. To obtain information about the power spectrum as a function of scale and time, wavelet coherence analysis \citep{Torrence1999,Grinsted2004} is used. Coherence analysis is able to quantify relationships between phases of two different signals, and measure whether the phase is constant (or locked) between two signals, as well as the difference in phase. This technique has previously been applied to different components of the magnetic field measured in a stream of fast solar wind at ion kinetic scales by \cite{Lion2016} and Perrone et al. 2017 \emph{accepted to APJ}. In the following sections we will present the data and discuss the methodology which will be followed by the results and a discussion. %Additionally we will seek to understand the source of the whistler waves and their interactions with electrons.

\section{Data/Methodology}

\begin{figure}
\centering
\includegraphics[angle=-90,width=0.8\textwidth]{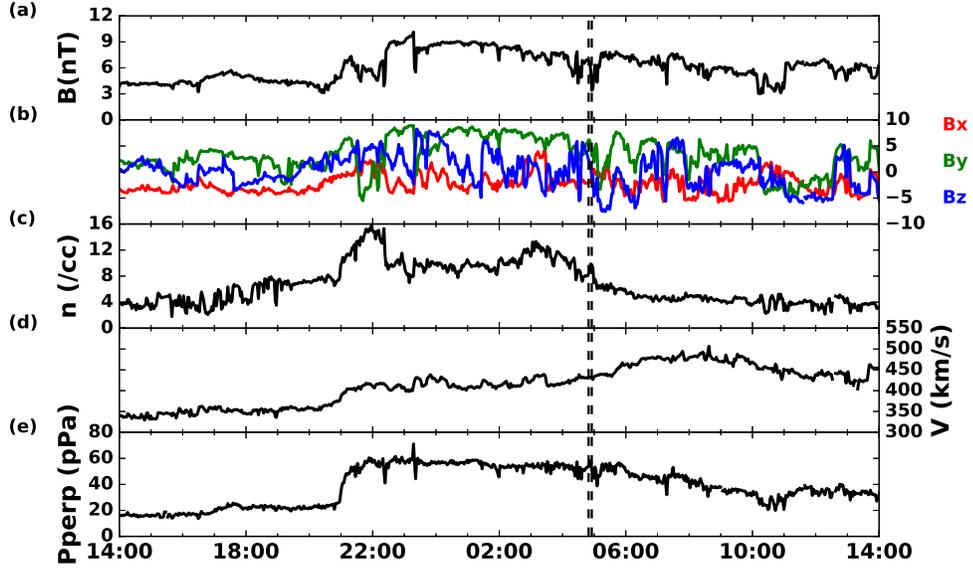}
\caption{Large scale plasma parameters of the solar wind plasma from {\it{Wind}} between the 2009-01-30 14:00 2009-01-31 14:00UT where the measurements have been propagated to the bow shock nose. The vertical lines denote the time of the {\it{Cluster}} data which we investigate. (a) shows the magnetic field magnitude, (b) shows the vector components, (c) shows the proton density, (d) shows the solar wind velocity, and (e) shows the total perpendicular pressure.}
\label{OmniData}
\end{figure}

In this paper an interval of slow solar wind, sampled by the {\it{Cluster}} mission during a burst mode interval occurring between 2009-01-31 04:49:23-04:54:23UT, is analyzed where the C2 spacecraft was located at $r_{GSE}=[12.3,11.8,-5.1] R_{E}$. Spectra from several sub-intervals of a larger interval of the same burst mode event were presented in \cite{Sahraoui2013,Sahraoui2013a}. For additional context measurements from the {\it{Wind}} spacecraft \citep{Acuna1995} of the solar wind upstream of Earth's environment during 24 hours around the interval of interest are plotted in Figure \ref{OmniData}. The vertical lines denote the corresponding data analyzed at {\it{Cluster}}. From the large scale data we see that our region of interest may be within a small scale Interplanetary Coronal Mass Ejection (ICME) or an interface between two streams of solar wind. The magnetic field measurements are obtained from the Magnetic Field Investigation \citep{Lepping1995} and are given in the Geocentric Solar Ecliptic (GSE) co-ordinate system where $z_{GSE}$, points towards the ecliptic north direction, $x_{GSE}$ points towards the Sun, while the final component is the vector product of the other two components $y_{GSE}=z_{GSE}\times x_{GSE}$. The plasma measurements are obtained from the Solar Wind Experiment \citep{Ogilvie1995}. At that time, {\it{Wind}} was located about 209 Re upstream of Earth. The data shown here have been shifted by 55 minutes to account for the propagation time from {\it{Wind}} to {\it{Cluster}}, as estimated from the average solar wind velocity during those 24 hours. Although {\it{Wind}} was at rather large distance from the Sun-Earth line ($Y_{GSE}$ $\sim -92$ Re), the large-scale structures it sampled should be representative of those arriving at Earth. 

Around 21:00 on 30 January 2009, the solar wind density (c) increases sharply. At the same time, the interplanetary magnetic field magnitude (a) and the solar wind velocity (d) also increase, though less remarkably. After the density jump, the magnetic field magnitude remains higher than in the preceding solar wind throughout the rest of the interval displayed here. This higher magnetic field strength is however not accompanied by smooth rotations of the IMF components, which would have hinted at the passage of a magnetic cloud \citep{Burlaga1981}. The structure starting at 21:00 may however correspond to a flank encounter of a rather slow and weak ICME, i.e. with no apparent flux rope structure. As this event takes place during solar minimum, the occurrence of such a weak ICME is likely. Panel (e) of Figure \ref{OmniData} displays the total perpendicular pressure, i.e. the sum of the plasma thermal pressure perpendicular to the magnetic field and of the magnetic pressure \citep[see][]{Jian2006}. The sharp increase of the total perpendicular pressure, followed by a gradual decrease over time, is rather similar to the Group 3 ICME presented in \cite{Jian2006}. The distinction between sheath part and magnetic obstacle is however not clear in our event because there is only a weak velocity jump and thus no sheath develops. Note also that the values of the total pressure are much lower in our case, because of the low IMF strength. The very slow decrease of the total pressure profile, in particular between 23:30 and 03:30, compared to the events of \cite{Jian2006}, could be due to the slightly faster solar wind which follows the structure and most likely compresses it, as shown by the high density between 21:00 and 05:00.
To summarize, the large scale solar wind data observations suggest that our interval of interest is embedded within a larger solar wind structure, which may be a very weak ICME caught up by a faster solar wind stream.

\begin{figure}
\centering
\includegraphics[angle=-90,width=0.9\textwidth]{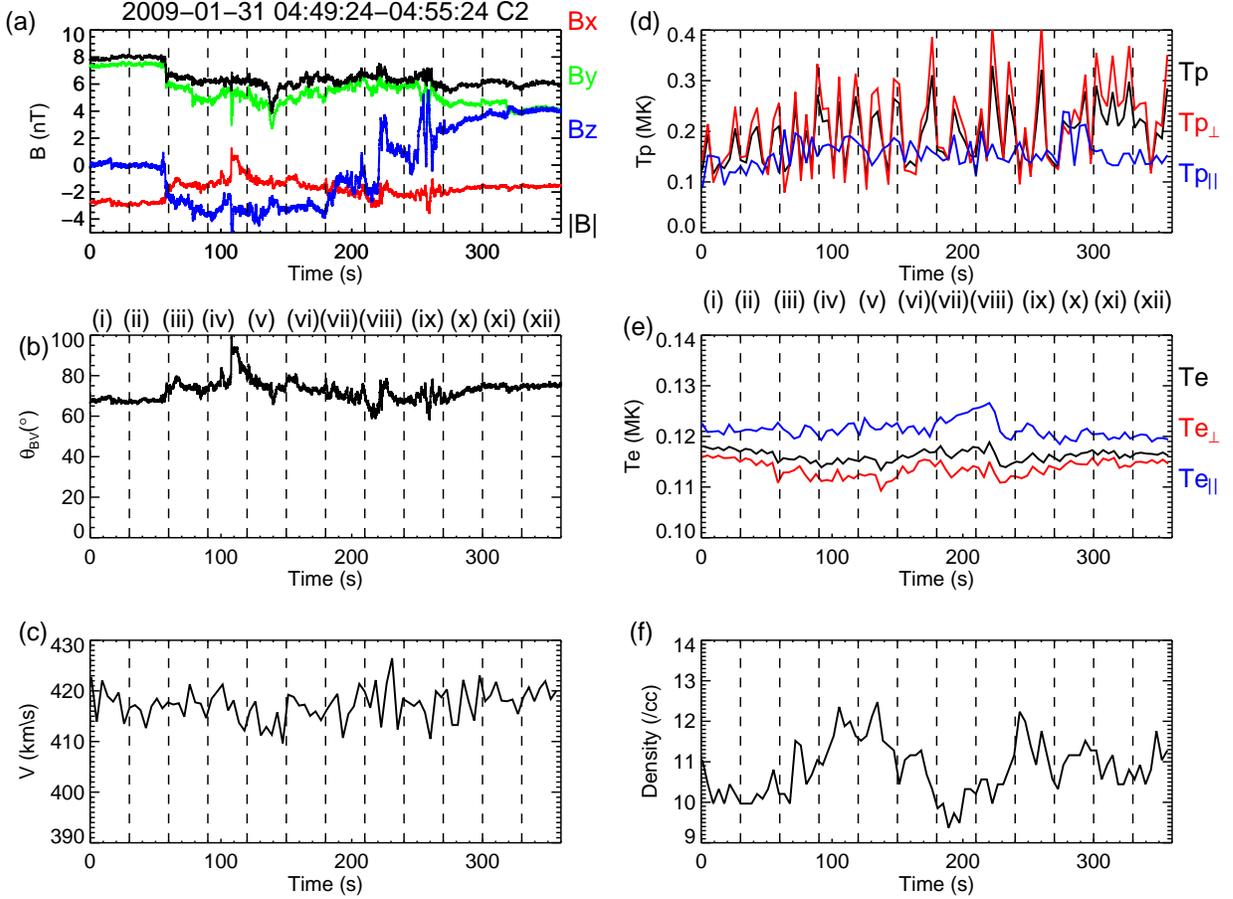}
\caption{(a) Large scale magnetic field data from {\it{Cluster}} 2 in GSE co-ordiantes. (b) Angle between the magnetic field and the solar wind bulk flow direction. (c) Proton speed and (d) Proton temperatures from CIS on {\it{Cluster}} 1. (e) Electron temperatures from {\it{Cluster}} 2. (f) Proton density from CIS on {\it{Cluster}} 1. Vertical lines denote time intervals which correspond to the spectra in figure \ref{MultiPS}.}
\label{RawData}
\end{figure}

Figure \ref{RawData} shows several of the plasma parameters for the shorter time interval of interest in the {\it{Cluster}} data. The large scale magnetic field time series obtained from the Fluxgate Magnetometer (FGM) instrument \citep{Balogh2001} on C2 is presented in Fig \ref{RawData}a and the angle between the magnetic field direction and the flow direction is presented in Fig \ref{RawData}b. This angle is larger than $60^{\circ}$ suggesting connection with the foreshock is unlikely. Moreover,  the electric field spectrogram from the WHISPER instrument \citep{Decreau1997}, and the ion distributions (not shown) from the {\it{Cluster}} ion spectrometer (CIS) \citep{Reme1997} on  C1 also suggest that there is no connection to the bow shock. The solar wind speed and the ion (from CIS on C1) and electron temperatures  \citep[from the plasma electron and current experiment (PEACE) on C2][]{Johnstone1997} are presented in Fig \ref{RawData}c, \ref{RawData}d and \ref{RawData}e respectively. Finally the proton density from CIS on C1 is shown in Fig \ref{RawData}f. The vertical lines show thirty second time intervals where we investigate the PSD. 

\begin{figure}
\centering
\includegraphics[width=0.90\textwidth]{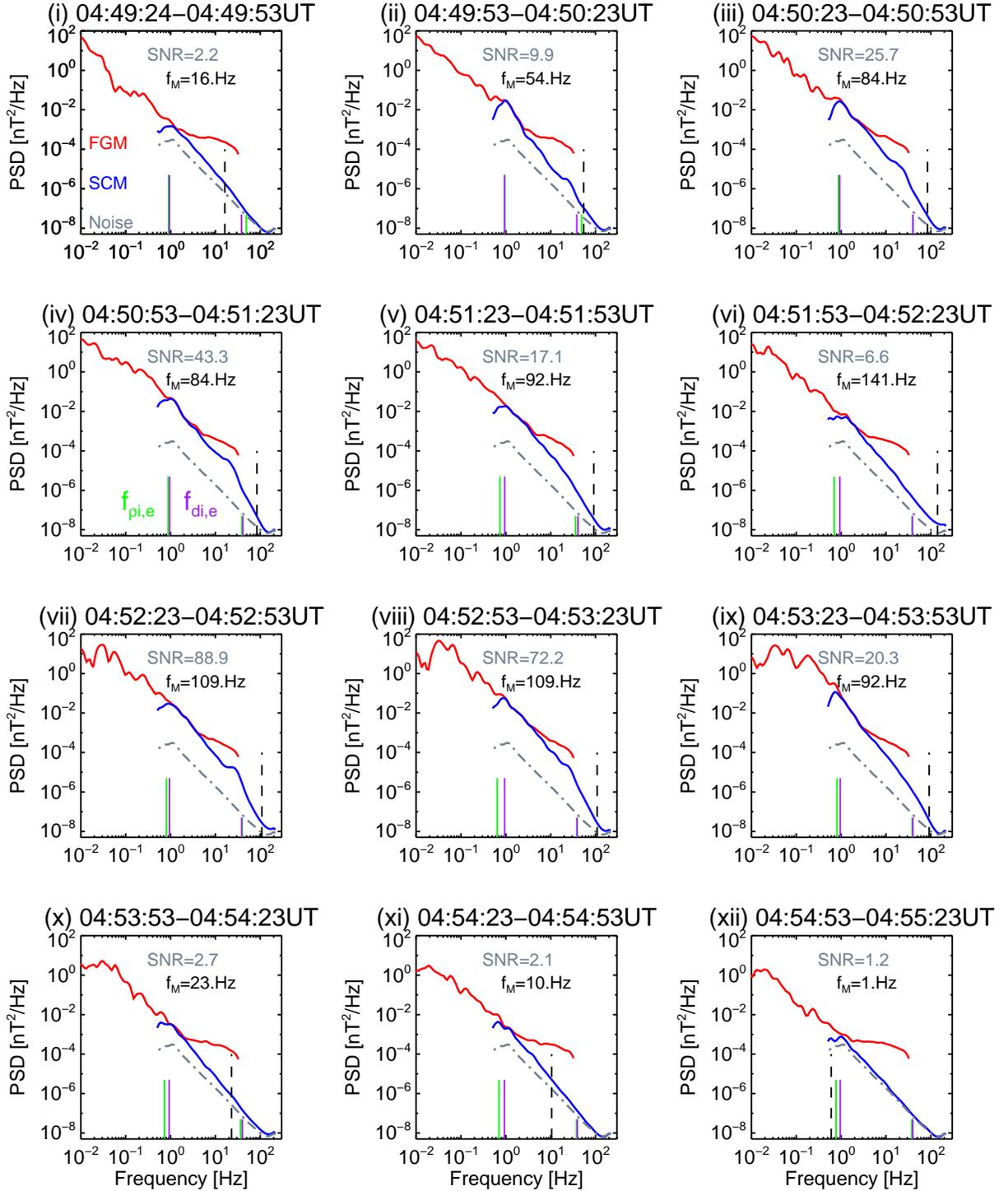}
\caption{Multiple power spectra which correspond to the regions between the dashed lines in Fig \ref{RawData}. Large scale spectra obtained from FGM and small scale fluctuations obtained from STAFF-SCM in Burst mode. The noise floor is estimated from an interval of Burst mode when {\it{Cluster}} was in the magnetic lobes. The Taylor shifted ion and electron inertial lengths $f_{\rho i,e}=V_{sw}/2\pi\rho_{i,e}$  and $f_{d i,e}=V_{sw}/2\pi d_{i,e}$ are indicated by the green and purple lines but are often very close and can lie on one another. The black dashed line indicates the maximum physical frequency described in the text. The electron cyclotron frequency is not displayed as it is larger than all of the largest physical frequencies $f_{ce}\sim 220Hz$.}
\label{MultiPS}
\end{figure}

At large scales the magnetic field data from the fluxgate magnetometer is used to probe frequencies between [0.001,1]Hz and data from the STAFF-SCM are then used to probe the small scales [1,180] Hz.  The PSD are calculated from the wavelet coefficients of the magnetic field time series \citep{Torrence1998}. To investigate the evolution of the PSD during the time of interest we divide the interval into 30 second sub-intervals and calculate the global wavelet spectra for each sub-interval in Fig. \ref{MultiPS}. This is done as to curtail any possible effects from averaging \citep{Sahraoui2013} and to limit the effect in the changing direction of the magnetic field when we perform wavelet coherency analysis later. 

The red lines in figure \ref{MultiPS} denote the power spectral density (PSD) estimated from the wavelet spectra of the FGM data. It should be noted that the wavelet transform is subject to edge effects due to the finite length of the time series, the so called cone of influence (COI), which is more marked at larger scales. To limit the influence of the COI in the FGM data we take the wavelet transform between 04:45:00-05:00:00 but construct the trace PSD by summing only the wavelet coefficients at the times of interest. The blue curves denote the search coil PSD at higher frequencies. For the SCM data the COI region is significant only in the low frequency range which have already been high pass filtered and thus we need not be concerned with these effects.  Generally between 0.4-1Hz both instruments show good agreement and similar PSDs, but at frequencies near $f_{sc}\ge 2$Hz the FGM instrument reaches the noise floor which can be seen as the flattening in the spectra starting at this frequency. The noise floor of the SCM is denoted by the grey dashed curves and the SNR at $f_{sc}=30$Hz is noted on each spectra and defined as:  $PSD_{\text{signal}}(30 \text{Hz})/PSD_{\text{noise}}(30 \text{Hz})$. The STAFF-SCM noise floor is estimated from an interval of magnetic field data when {\it{Cluster}} were in the magnetic lobe of Earth's magnetosphere. This is the same as that used by \cite{Kiyani2009a,Kiyani2013} and was verified to contain Gaussian fluctuations consistent with random noise.  The SNR at 30Hz was chosen as this is the approximate frequency where we see enhancements in some spectra in Figure \ref{MultiPS}. A maximum physical frequency $f_{\text{M}}$ is also indicated and defined as the frequency where the SNR=3. When SNR is below 3 then the signal is no longer physical and is contaminated by noise \citep{Alexandrova2010a}. The SNR is sufficiently large for spectra ii-ix in figure \ref{MultiPS} such that study of the region near the shifted electron characteristic scales can be performed with no contamination from noise. The values of the various proton and electron shifted scales are also indicated by the green and purple lines which do not vary much in the interval, and are often close to one another as the plasma beta for protons and electrons (ratio of thermal pressure to magnetic pressure) are close to one $\beta_{p}=1.84\pm 0.72$, $\beta_{e}=1.01 \pm 0.23$.

It is clear from Figure \ref{MultiPS} that the power spectrum at electron scales can change quickly even within a short time range. The spectra also show a variety of different spectral shapes with some showing a clear breakpoint (iii,viii), some without a distinct break (vi,x,xi), others showing an exponential cutoff (v,ix) and some showing large enhancements near the Taylor shifted electron scales (ii,iv,vii). To further investigate the spectral shapes seen, wavelet coherence is used \citep{Torrence1999,Grinsted2004,Lion2016}.  The wavelet coherence and phase difference angle between a pair of signals are defined in Eqs. \ref{coherence},\ref{phaseangle} as in \cite{Torrence1999,Lion2016}. 

\begin{equation}
R^{2}_{ij}(f, t) = \frac{| S(fW_{i}(f, t)W_{j}^{*}(f, t))|^{2}}{S(f| W_{i}(f, t)|^{2}) \cdot S(f|W_{j}(f, t)|^{2})}
\label{coherence}
\end{equation}

\begin{equation}
\phi_{ij}(f,t)= \tan^{-1}\left(\frac{\Im\left({S(fW_{i}(f, t)W_{j}^{*} (f, t))}\right)}{\Re{(S(fW_{i}(f, t)W_{j} (f, t)^{*}))}}\right)
\label{phaseangle}
\end{equation}

Here $f$ denotes the frequency (inverse time scale), $t$ denotes the time, $W$ denotes the complex Morlet wavelet coefficients. The asterisk denotes the complex conjugate, $S$ denotes a scale dependent smoothing operator and Fraktur fonts denoting real and imaginary parts of the complex wavelet coefficients. The subscripts $i$ and $j$ denote the different signals, for our purpose these two signals will be two components of the magnetic field.

%in the plane perpendicular to the local mean (30 second average) magnetic field direction. While other components such as the magnetic field time series in the direction parallel to the magnetic field direction can also be analyzed with respect to one of the perpendicular components at these scales very little coherence was found at the scale of interest when compared to the two perpendicular components. Therefore we will restrict ourselves to the analysis of the two perpendicular components.}  

The coherence gives a measure of the phase relation of two different signals and has values between 0 and 1, where 1 denotes full coherence where the signals exhibit a constant phase relation (or phase locking), while a value of 0 denotes no phase relation between the two signals. In the study of \cite{Lion2016} the fast solar wind power spectrum was found to consist of incoherent fluctuations (making $\sim 40\%$ of the interval), with regions of strong coherence in the plane perpendicular to the mean magnetic field consistent with ion cyclotron waves (contributing to $\sim20\%$ of the interval), and the remainder of the fluctuations in the plane subtended by $\mathbf{B_{0}}$ and $\mathbf{v_{sw}}$ were found to be characteristic of coherent structures such as Alfv\'enic vortices \citep{Alexandrova2006,Alexandrova2008a,Roberts2016} and current sheets. 

At electron scales strong coherence for a small range of frequencies in the plane perpendicular to the magnetic field in tandem with a phase difference of $+90^{\circ}$ would indicate parallel propagating right hand polarized magnetosonic/whistler waves \citep{Lacombe2014}. Coherent structures exhibit coherence over multiple scales simultaneously \citep{Frisch1995,Lion2016}. Following \cite{Lion2016} we can define a threshold for coherency at a certain scale to separate the power spectra into regions with fluctuations characteristic of waves, those characteristic of coherent structures as well as an incoherent component. To define the threshold and to obtain statistical confidence that two signals are coherent the following method is used, having first been presented for application to solar wind spectra in \cite{Lion2016}. The data are Fourier transformed and the phases of the signal are randomized in Fourier space \citep{Koga2003,Hada2003}. Then the inverse Fourier transform is performed on the random phase signal, which results in a signal which has the same power spectrum as the initial signal while the phase information is random. Using this approach we can investigate the coherence of the two random phase signals. At the scale of interest we take a mean coherence of 100 signals, and the standard deviation is used to define our threshold as a function of time $R_{ij}^{\text{threshold}}(t)=\bar{R_{ij}}(t)+2\sigma_R (t)$ as was done in  \cite{Lion2016}. To investigate electron scales we chose 30Hz as the scale of interest where the signal to noise is sufficiently large in spectra ii-xi for further investigation, and where a large bump can be seen in spectra ii,iii,iv vii and viii.           

\section{Results}

\begin{figure}
\centering
\includegraphics[width=0.95\textwidth]{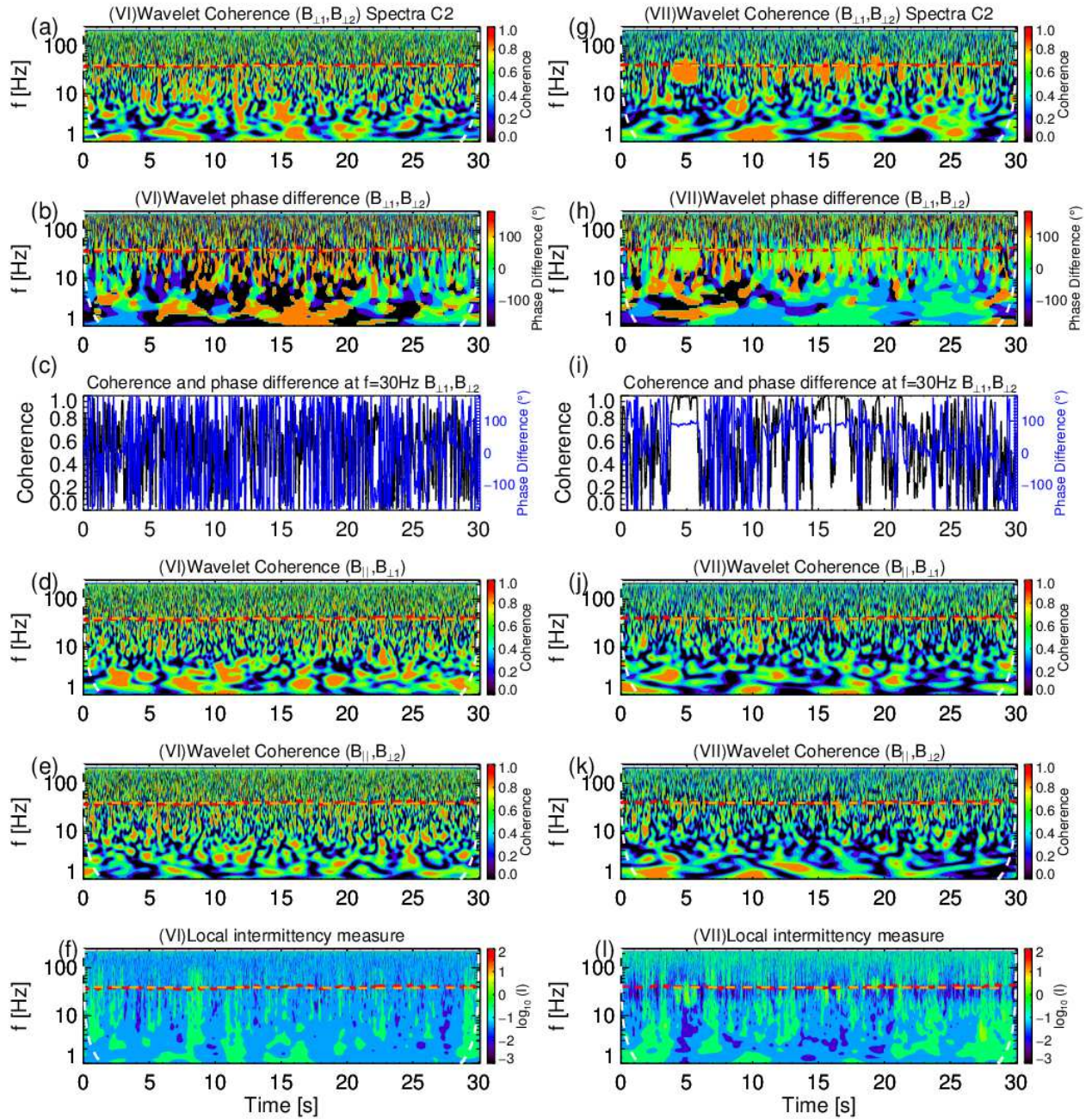}%\hspace*{0.7cm}
\caption{Coherence, phase and local intermittency measure for spectra vi and vii. Panels a,g show the wavelet coherence as a function of time and scale for the magnetic field fluctuations in the two perpendicular components to the large scale magnetic field direction. Panels b,h show the wavelet phase between the two components. \textbf{Panels c,i show a one dimensional cut of the coherence and phase difference at the scale of interest $f=30$Hz for the two perpendicular components (corresponding to the two panels above, a,g and b,h).}  Panels d,j and e,k show the wavelet coherence between two different components of the magnetic field one parallel and one perpendicular. Panels e,j show the trace local intermittency measure. Orange and red lines denote the Taylor shifted electron inertial and Larmor frequencies which practically lie on one another. White lines denote the cone of influence region of the wavelet transform.}
\label{wave1}
\end{figure}

Figures \ref{wave1}a,f show the wavelet coherence between the two components perpendicular to the mean magnetic field direction, with time intervals corresponding to the global spectra vi-vii in Figure \ref{MultiPS}. Panels c,d,h,i show the coherence between different pairs of signals where we define the co-ordinate system as $\mathbf{e}_{\parallel}=\mathbf{B_{0}}/|\mathbf{B_{0}}|$, the mean magnetic field is defined as a local 30 second mean from FGM during the time the spectra is taken, as to avoid any complications due to the use of a global mean. The two perpendicular components are defined as $\mathbf{e}_{\perp 1}=\mathbf{e}_{\parallel} \times \mathbf{V_{sw}}/|\mathbf{V_{sw}}|$ and $\mathbf{e}_{\perp 2}=\mathbf{e}_{\parallel}\times \mathbf{e}_{\perp 1}$. In contrast to the coherence of fluctuations investigated at ion kinetic scales where strong coherence was seen both in $\mathbf{e}_{\perp 1}-\mathbf{e}_{\perp 2}$ (interpreted as ion cyclotron waves) and $\mathbf{e}_{\parallel}-\mathbf{e}_{\perp 2}$ (interpreted as Alfv\'enic vortices and current sheets) there is typically very little coherence between pairs of signals other than $\mathbf{e}_{\perp 1}-\mathbf{e}_{\perp 2}$. Moreover, when there is strong coherence in the parallel/perpendicular components it occurs simultaneously with strong coherence in the two perpendicular components. The phase difference for the two perpendicular components is given in panels b,g. Finally, the wavelet trace power $W_{\text{trace}}$ normalized to the mean at each scale (defined in Eq. \ref{Lim1})  which is often termed the local intermittency measure $I$ \citep{Farge1992} in panels e,j respectively allows for clear identification of intermittent events in time and frequency such as coherent structures/discontinuities or sporadic wave activity. A horizontal dashed line is also shown and marks the Taylor shifted electron Larmor and inertial frequencies (which are very close to one another), which are calculated each four seconds with the spin resolution magnetic field and electron plasma data.

\begin{equation}
I(f,t)=\frac{| W_{\text{trace}}(f,t) |^2}{\langle | W_{\text{trace}}(f,t) |^2 \rangle}
\label{Lim1}
\end{equation}

We will now focus on the analysis of $\mathbf{e}_{\perp 1}-\mathbf{e}_{\perp 2}$ pair of signals as the coherence here dominates other components in terms of the value of coherence and the duration. Coherence maps, phase difference, and the local intermittencey measure for the two perpendicular components are provided for intervals viii and ix in Figure \ref{wave2}. We chose these four spectra vi-ix for further analysis since they represent a wide diversity of different morphologies and are close to each other in time so they have similar plasma parameters as well as having sufficient SNR.  The three wavelet products show several features. We will now discuss their relation to the global spectra presented in Figure \ref{MultiPS}.

Spectrum vi (Figure \ref{wave1}a,b,c,d,e,f) does not show any clear spectral break or significant steepening at electron scales. The coherence analysis shows the presence of several elongated regions of high coherence across several scales up to around 20 Hz which are characteristic of coherent structures. Spectrum vii (Figure \ref{wave1}f,g,h,i,j) is perhaps the most striking showing a large bump at the electron characteristic scales. For this spectrum, wavelet analysis reveals energetic intervals of high coherency at the frequency of the bump in Figure \ref{MultiPS} vii. The most notable interval occurs between 4-6s, although there are numerous shorter bursts later between 11 and 20 seconds. These regions of high energy and coherence are also associated with phase of $+90^{\circ}$ (right hand polarization) consistent with parallel propagating whistler waves. This is shown more clearly in the one dimensional cuts at 30Hz in Fig \ref{wave1}i, where the left axis and black lines denote the coherency and the right axis and blue lines denote the phase difference.  The contrast between panels c and i is rather striking. Whereas no clear highly coherent regions can be seen in panel c, panel i shows regions where the coherence is close to one with a corresponding phase difference between the components of $+90^{\circ}$ (right hand polarization). These properties are consistent with whistler waves propagating along the magnetic field direction.

\begin{figure}
\centering
\includegraphics[width=0.95\textwidth]{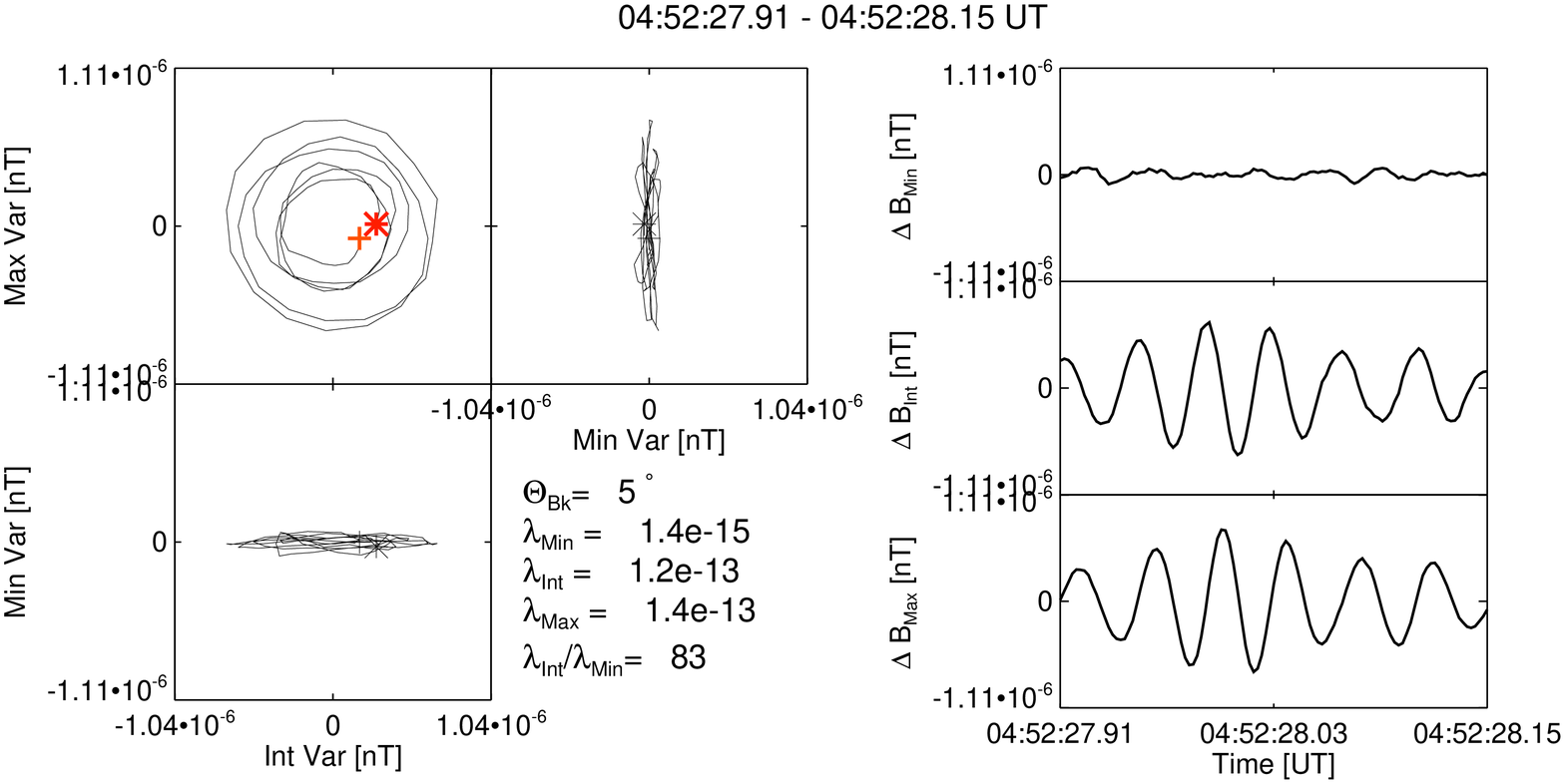}
\caption{Minimum variance analysis showing the hodographs of the fluctuations in the maximum intermediate and minimum planes (left) for the most energetic wavepacket between 4-6 seconds in interval vii. The corresponding waveforms are presented in the three panels (right).}
\label{hod1}
\end{figure}

\begin{figure}
\centering
\includegraphics[width=0.95\textwidth]{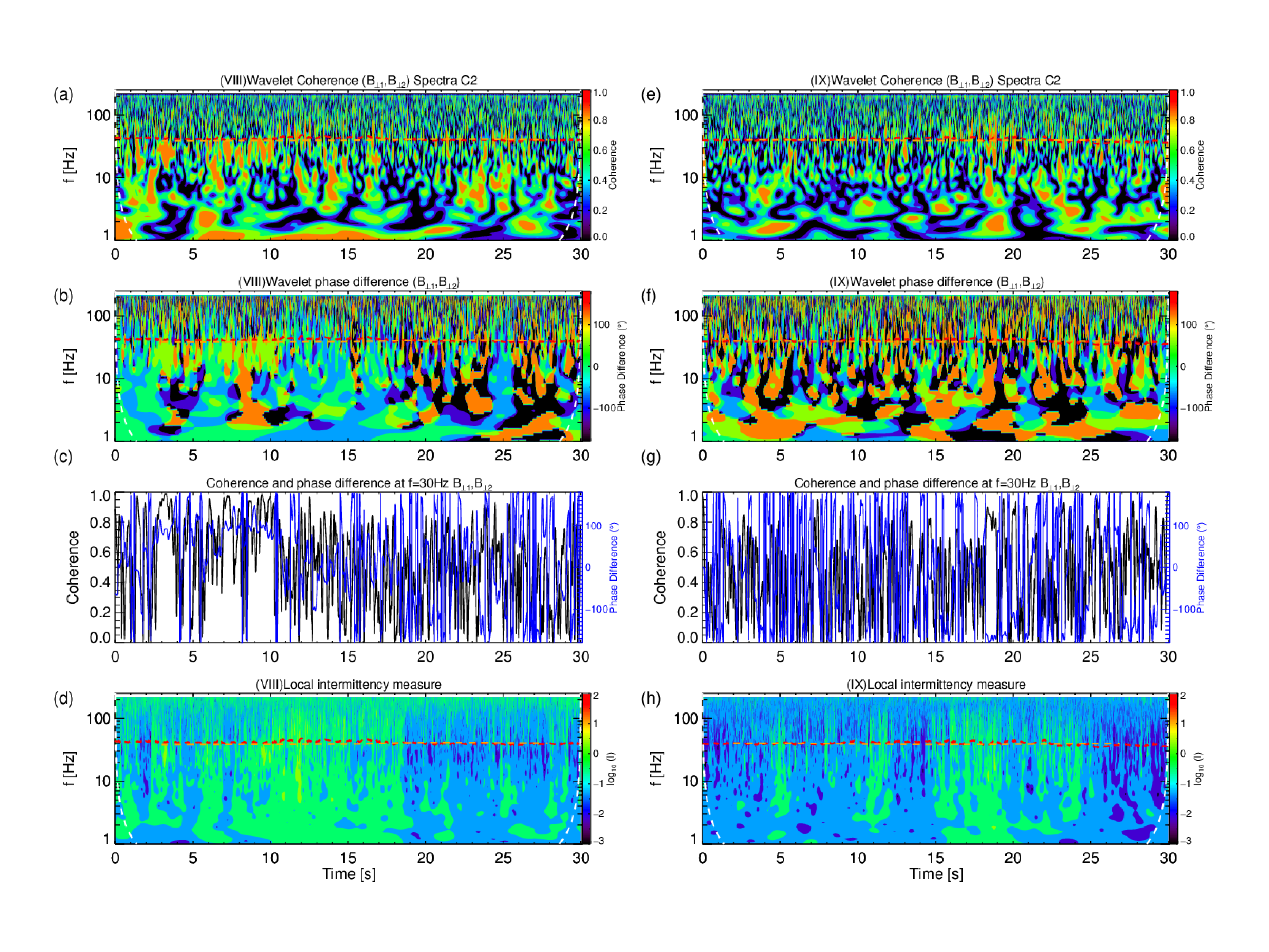}%\hspace*{0.7cm}
\caption{Wavelet coherence for two magnetic field components perpendicular to the mean magnetic field (a,e), the phase difference (b,f), the 1D cut through $f=30$Hz (c,g) and the local intermittency measure (d,h)  for spectra viii and ix}
\label{wave2}
\end{figure}

To confirm that these fluctuations are indeed parallel propagating whistler waves minimum variance analysis \citep{Sonnerup1998} is performed between 04:52:26.85-04:52:27.08 (the region of large coherence and a $+90^{\circ}$ phase difference of the two signals between seconds 4 and 6 in Figure \ref{wave1}de). The eigenvalues retrieved from the analysis $(\lambda_{\text{Max}},\lambda_{\text{Int}},\lambda_{\text{Min}})/\lambda_{\text{Max}}=(1,0.85,0.01)$ and the ratio of the intermediate to minimum eigenvalues is large (82) giving confidence that the analysis is valid. The hodogram of the fluctuations is shown in Figure \ref{hod1}, and demonstrates that the fluctuations are right hand circularly polarized. Moreover, the wavevector $\mathbf{k}$ makes a small angle ($\theta_{\mathbf{kB}}=5^{\circ}$) with the mean magnetic field direction confirming that these fluctuations are indeed parallel propagating whistler wave trains.  

Spectrum viii (Figure \ref{wave2}a,b,c)  shows similar features as those found in spectrum vii that are characteristic of whistler waves between seconds 3 and 10. There are a number of other features present, periods more characteristic of coherent structures at times 1s, 12s, 16s and 24s (Figure \ref{wave2}abc). Finally Spectrum ix (Figure \ref{wave2} d,e,f) shows only the presence of structures. However, in contrast to spectrum vi the coherence in this case extends up to and some even exceeding the electron characteristic scales. In spectrum ix, there are no strong emissions concentrated near 30Hz as in spectra vii and viii, however there are several small regions where coherency exists over a large range of frequencies near 5s, 12s, 23s and most prominently at around 19s but there are no clear signals of whistler emission. In these four different cases the coherency reveals the signals contain different fluctuation types.

	To investigate the influence these fluctuations have on the morphology of the spectrum we apply the method of \cite{Lion2016} described in section 2 to select wavelet coefficients which are above $R_{ij}^{\text{threshold}}(t)$ which make up coherent times and those which are below the threshold making the incoherent times. We only consider the coherence of the two perpendicular components with respect to the 30s mean magnetic field. As we have mentioned above, the coherence between 2 perpendicular components dominates the other pairs (shown in Figures \ref{wave1}d,ej,k.) especially in duration for coherent events around 30Hz. Here we use the term `incoherent' to refer to the times when the signal does not reach the coherence threshold for the two perpendicular components. However, the same times may in reality be coherent between two different pairs of signals; for example a pair of signals might exhibit strong coherence for the two directions perpendicular to the wavevector $\mathbf{k}$ i.e. for a perpendicular propagating plasma wave. 
	
\begin{figure}
\includegraphics[width=0.95\textwidth]{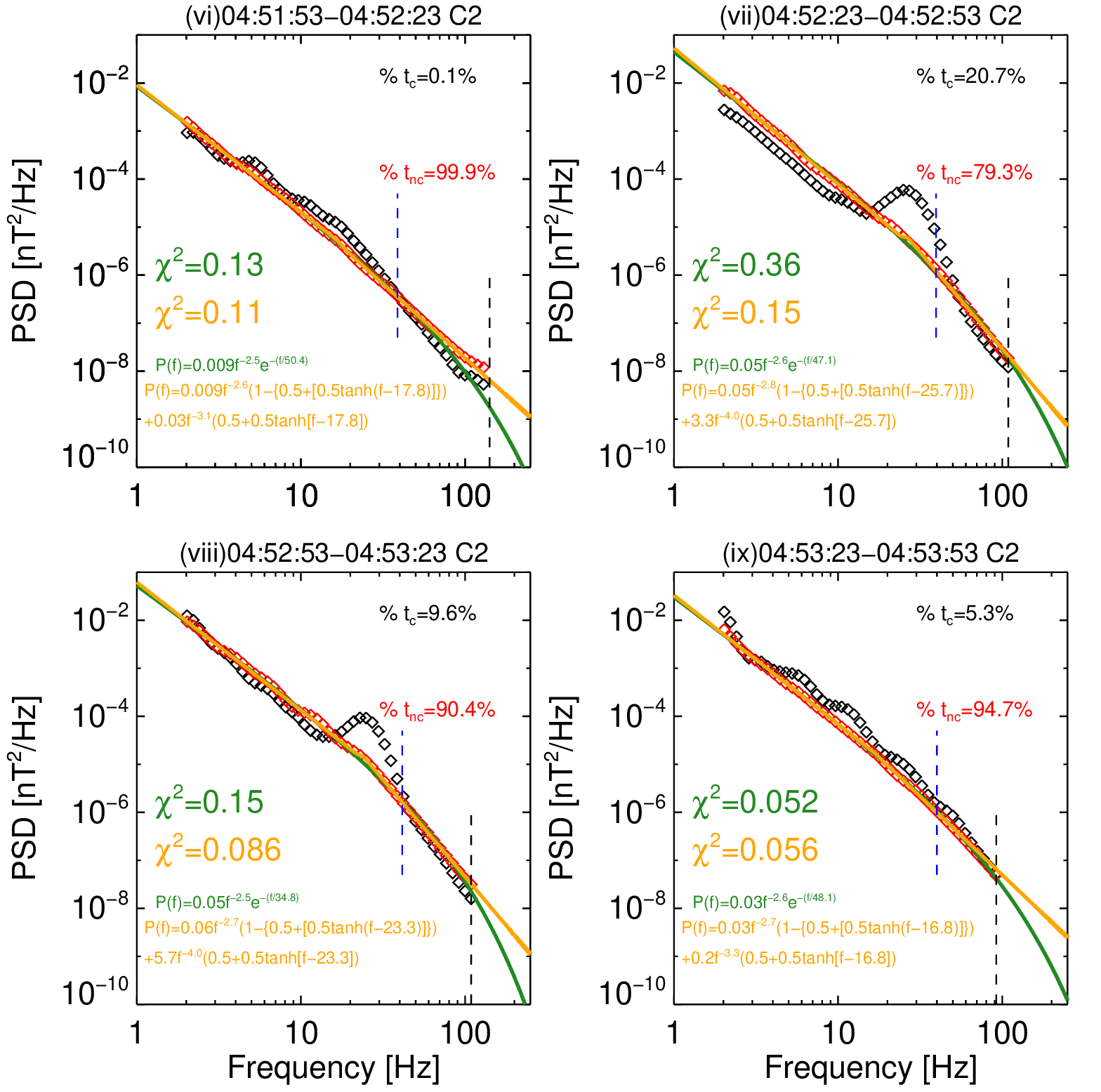}
\caption{Spectra which have been averaged according to their coherence properties. The incoherent component has been fitted with the two models between $2$Hz and the maximal physical frequency $f_{M}$ which is denoted by a black dashed line. The blue dashed line denotes the Taylor shifted electron gyroradius which is approximately equal to the shifted electron inertial length. Black spectra denote the regions which are coherent and above the coherency threshold, meanwhile red denote the incoherent components of the spectra which did not meet the threshold criteria. The percentage of time where the spectra are coherent $\% t_{c}$ and not coherent $\% t_{nc}$ are indicated on the figures.}
\label{Threshold}
\end{figure}

Figures \ref{Threshold}vi-ix show spectra corresponding to the same Roman numerals as in Figure \ref{MultiPS}, where the black diamonds denote the spectra integrated over coherent times and the red diamonds denote the spectra integrated over incoherent times. The technique succeeds in identifying regions where there are whistler waves (the coherent regions) which are shown in black in Figure \ref{Threshold}. The percentage of time in each interval where it is coherent and incoherent is indicated on each panel. In spectrum vii the coherent fluctuations make a significant part of the global power despite only being present during $20\%$ of the time. Interestingly spectrum viii shows clear whistler waves also, however since they are not as energetic or numerous they cause a clear break in the spectrum rather than an enhancement. In spectrum ix the coherent component of the fluctuations shows something different with power across a broader range of scales stretching from a few Hz to 30Hz suggesting that these are not whistler waves (concentrated at 30Hz) but coherent structures which exhibit coherence over many scales.

Two models are fitted to the incoherent components of the spectra between $2$ and $f_{\text{M}}$Hz using a nonlinear least squares method \citep{Markwardt2009}. The first model is the exponential model given by Eq. \ref{exp} \citep{Alexandrova2012}, which consists of a power law and an exponential steepening and is shown in green. The second is the break model (Eq \ref{break}), which consists of two power law slopes separated by a spectral break \citep{Alexandrova2012,Sahraoui2013} and is shown in orange. To avoid numerical difficulties fitting the Heaviside function which is not differentiable at the break, we use a hyperbolic tangent function which approximates the Heaviside function but remains differentiable at the break and is given in Eq \ref{HypTan}.  

\begin{equation}
PSD(f)=A_{1}f^{- \alpha_{1}}(1-{0.5+[0.5\tanh(f-f_{b})]})+A_{2}f^{-  \alpha_{2}}(0.5+0.5\tanh[f-f_{b}])
\label{HypTan}
\end{equation}

It is clear that both models fit the data well with low values of the $\chi^{2}$ statistic which is defined as:

\begin{equation}
\chi^{2}=\Sigma_{f} \left(\frac{\text{PSD}(f)-\text{Model}(f)}{\sigma(f)}\right)^2
\label{chi}
\end{equation}

where $\text{Model}(f)$ is the model used and the PSD$(f)$ and the error $\sigma (f)$ are defined in Eq \ref{PSD1},\ref{PSD2} and the angle brackets denote an average over the incoherent times:

\begin{equation}
PSD(f)=2 dt \langle |W(f,t)|^{2} \rangle_{t_{\text{inc}}}
\label{PSD1}
\end{equation}

\begin{equation}
\sigma(f)=2 dt \sqrt{\frac{1}{N_{\text{inc}}}\sum_{t=t_{\text{inc}}}\left(|W(f,t)|^2 -\langle |W(f,t)|^{2} \rangle \right)^{2}}
\label{PSD2}
\end{equation}

It is important to note that this definition of $\chi^{2}$ does not account for the number of free parameters. Both models perform comparably, however the break model has more free parameters and would be expected to have a better fit generally. This is not the case for all spectra, the incoherent component of spectrum ix shows better agreement with the exponential model. However, it must be stressed that the regions where the fitting is the poorest are at the smallest scales where the amplitudes of the fluctuations are smaller and the SNR is lower making the measurement a challenging one.

\section{Discussion and Conclusion}
%\subsection{Influence on the magnetic field power spectrum}

The analysis has shown that the strongest coherence is between the two components perpendicular to the magnetic field. The coherence results here contrast strongly with the interval of fast wind at ion kinetic scales studied by \cite{Lion2016} where the coherence was most prevalent in the $\mathbf{e}_{\parallel}-\mathbf{e}_{\perp 2}$ components. It is unclear what is the reason for this difference, perhaps there are fewer coherent structures at these scales. Alternatively at smaller scales the spacecraft pass through fewer structures due to their smaller size, however coherent structures are observed but in the pair of components perpendicular to the mean magnetic field direction therefore we discard this possibility. The final possibility is that the coherent structures at these scales may be less compressible than those at ion kinetic scales. This would result in the $\mathbf{e_{\parallel}}$ component possibly exhibiting less coherence with respect to the other components than at ion kinetic scales where coherent structures are mostly compressible vortices (\cite{Perrone2016}, Perrone et al. 2017 \emph{accepted to APJ}).

\begin{figure}
\includegraphics[width=0.45\textwidth]{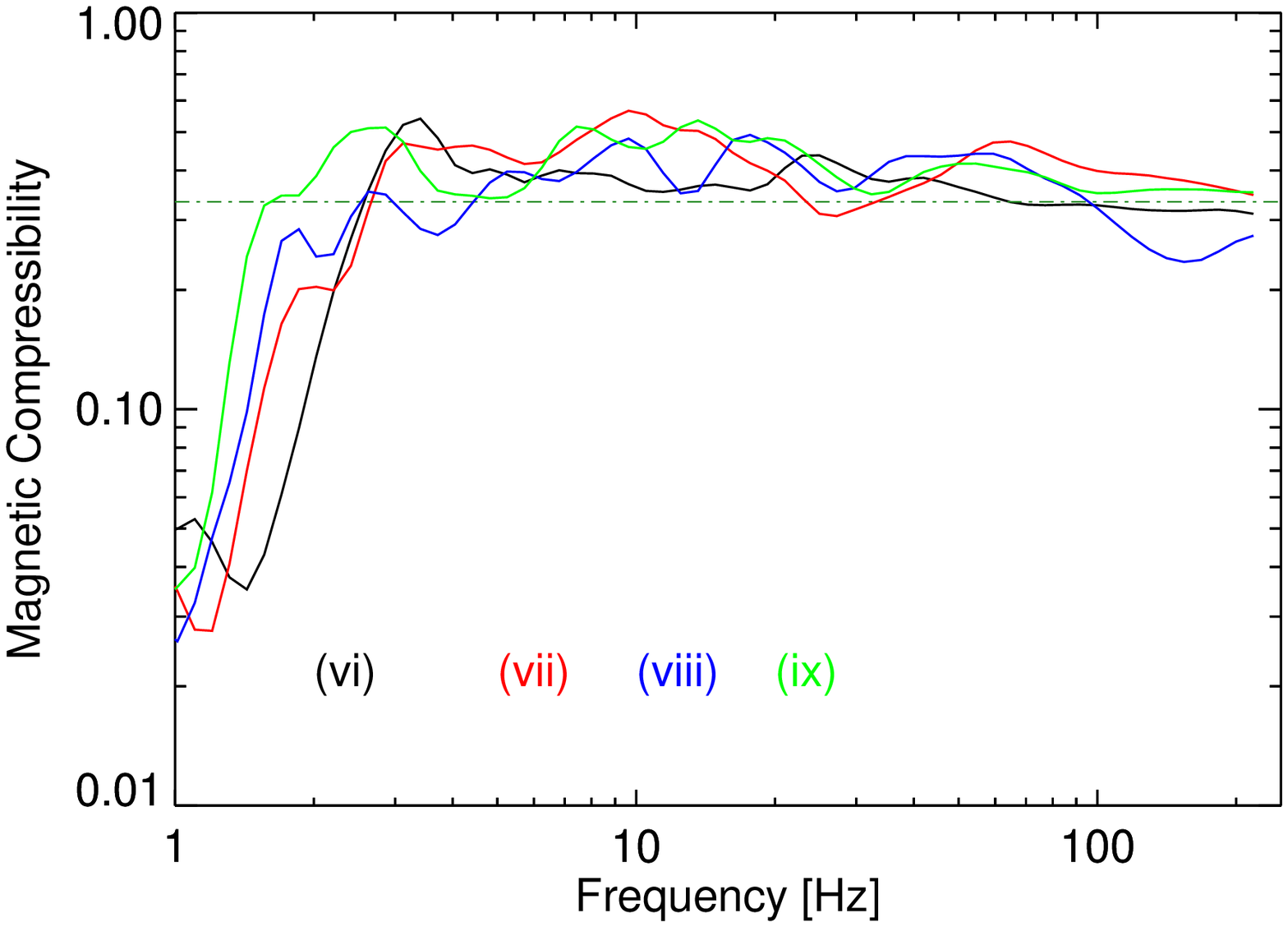}
\caption{Magnetic compressibility from the magnetic field fluctuations measured by STAFF for the four intervals of interest. The dot-dashed dark green line denotes isotropy at $C_{\parallel}=1/3$.}
\label{Comp}
\end{figure}

However typically at electron scales there is an increase in compressibility (\cite{Kiyani2013},Lacombe et al. 2017 \emph{accepted to APJ}). Figure \ref{Comp} shows the compressibility for the four intervals of interest here, and in these cases the compressibility increases sharply at electron scales. We define the magnetic compressibility as $C_{\parallel}(f)={\langle  |W_{|B
|}(f,t) |^2\rangle_{t}}/{\langle |W_{\text{trace}}(f,t) |}^2\rangle_{t}$ and a line at $C_{\parallel}=1/3$ denotes isotropy. If the compressibility increases at these scales while simultaneously there are fewer compressible coherent structures, what causes the increase in compressibility at electron scales? These may be due to quasi perpendicular propagating waves such as the KAW/KSW or ion Bernstein waves, which will be discussed later in this section in relation to the spectral slopes of the incoherent component.

It is clear that parallel whistler waves have a significant role in the shaping of the power spectrum at electron scales but are only present for a very narrow frequency range ($\sim 20-35$ Hz). In the more extreme cases as in Figure \ref{MultiPS},iii,vii, they can produce an enhancement at electron scales. When they make up a smaller fraction of the global PSD, they can result in a clear spectral break as in Figure \ref{MultiPS} viii. When the total spectra (where the wavelet coefficients are summed over the entire time interval) are fitted, spectrum viii shows better agreement with the break model $[\chi^2_{\text{break}},\chi^2_{\text{exp}}]=[0.16,0.3]$ and the coherence analysis shows $10\%$ of the time interval consist of whistler waves.  Therefore a spectral break or a knee represent the same fundamental process, and how these manifest in the global spectrum is a question of the amplitude and duration of the parallel propagating whistler waves.  

It is unclear what the source of the whistler waves is, some possibilities include a plasma instability such as the whistler anisotropy instability, the electron heat flux instability or a two stream instability. The other possibility is that the waves are related to coherent structures, either that they can excite waves directly or cause deformations in the electron velocity distribution which then emits waves to stabilize. In sub-intervals vii and viii where whistlers are observed, strong discontinuities are also observed in the $B_{z}$ component of the large scale magnetic field. The final possibility for the generation of whistler waves is that they are related to the large scale structure of the plasma, and a result of the interaction of a weak ICME with the ambient solar wind, or a stream interaction region as mentioned in section 2. The question of the generation of such short lived whistler emission will be explored further in a future work.

The solar wind stream in particular seems like an atypical case. However we may ask whether such whistler wave emissions are typical in `quiet' streams of solar wind. The statistical study of \cite{Lacombe2014} showed that $10\%$ of their data contained energetic whistler waves. The events containing whistler waves were predominantly classed as slow solar wind ($v_{sw}< 500$kms$^{-1}$) where the thermal pressure was low ($p_{th}<0.04$ nPa). Whereas in fast solar wind streams with high pressure \cite{Lacombe2014} hypothesised that the background turbulence was too energetic and any whistler waves would be too weak to manifest themselves in the power spectrum. To explore whether variability in the solar wind conditions lead to the presence of whistler waves a larger study is required which is outside the scope of the current work. Wavelet coherence would be able to identify such weak whistler waves in the data even if the background turbulence is much stronger since it is a measure of the phase relation between two components and the power is unimportant.

%This causes a spectrum with an exponential form e.g. Fig \ref{MultiPS} ix. This is unlikely to be an averaging effect as discussed by \cite{Sahraoui2013} since during these short 30s intervals where neither the solar wind velocity (Fig \ref{RawData} c), magnetic field strength (Fig \ref{RawData} a) or electron temperatures (Fig \ref{RawData} d) change appreciably in the interval, \textcolor{blue}{less so in each of the sub-intervals.} %Additionally the does not show any large changes as a function of time. 

Contrary to whistler waves which cause enhancements to appear in a narrow range in the spectrum, coherent structures contribute to the total power over a larger frequency range ($\sim 1-50 $Hz). The presence of coherent structures at electron scales is demonstrated here to coincide with a spectrum with a clear exponential morphology in spectrum ix (where the total spectrum has the following values of $\chi^2$ $[\chi^2_{\text{break}},\chi^2_{\text{exp}}]=[0.14,0.078]$) which is characteristic of strong turbulence. It is interesting to note that during interval ix where coherent structures are observed at electron scales in the wavelet coherence in the large scale data from FGM there is a clear coherent structure lasting for roughly ten seconds in the Bz component of Fig 2. This emphasizes the need to consider many scales simultaneously for the analysis of turbulence, and simply focusing on one scale gives an incomplete description.

The coherence analysis performed is very effective at isolating coherent events, whether they are waves or coherent structures. However, the dominant component in the spectra is the incoherent one which has no strong coherence between two perpendicular components. It makes up the largest fraction of the turbulent power in any of the intervals presented here. This component does not vary much between each of our cases. The shape of the total power spectrum at electron scales can be significantly influenced by the coherent components with the most dramatic effect coming from parallel whistler waves which influence the spectrum over a short frequency range in contrast to coherent structures which have power over a large range of frequencies.  Is it possible that the component without strong coherence between perpendicular components represent a universal spectrum of turbulent fluctuations \citep{Alexandrova2009,Alexandrova2012}? The intervals of \cite{Alexandrova2009,Alexandrova2012} were chosen so that they contain no whistler waves, thus they would only contain a background component and coherent structures and showed better agreement with the exponential model. Thus the shape of the total spectrum is composed of a background component, a component of coherent structures (which both have exponential spectral shapes) while whistler waves give characteristic enhancements at their frequency which coincides in the solar wind to the Taylor shifted spatial electron scales leading to a morphology for the total spectrum which agrees better with the break model and in extreme cases can lead to a spectral knee. 

In the four individual cases studied in Figure \ref{Threshold} the spectra of the incoherent fluctuations are similar in all cases and can be modeled equally well by either the break or the exponential model, however the exponential model has the advantage of having fewer free parameters. The region where these two models begin to diverge significantly ($f\geq 100$ Hz) happens to be where noise becomes a significant factor. Future observations of the high frequency range of the solar wind turbulent power spectrum are required with a much higher sensitivity to determine which model agrees better with the data.

The spectral index which are fitted to the data vary between $\alpha \in [-2.5,-2.6]$,$\alpha_{1} \in[-2.6,-2.8]$ above 30Hz and $\alpha_{2} \in [-3.1,4.0]$.  These values are consistent with the majority of intervals of the trace magnetic spectra surveyed previously \citep{Alexandrova2009,Alexandrova2012,Sahraoui2013}. The Taylor shifted electron gyroradius and inertial scales are all similar $f_{\rho_{e}}\sim f_{d_{e}} \sim 40$ Hz and are indicated by the blue dashed line. The break and cutoff frequencies vary as $f_{b} \in [17.8,25.7]$Hz and $f_{c} \in [47.1,50.8]$Hz, with the cutoff frequencies showing better agreement with the shifted electron scales. An important question is: what is the nature of the dominant incoherent spectrum?

Several possibilities exist; the incoherent fluctuations may be fluctuations which occur only in a single component, which would not be recovered by coherence methods as they quantify relationships between pairs of signals. Another possibility is that these fluctuations could be due to quasi-perpendicular waves as discussed previously in relation to the compressibility, candidates include KAWs \citep[e.g.][]{Howes2008b,Chen2013}, ion Bernstein/whistler waves \citep[e.g.][]{Coroniti1982,Stawicki2001}. KAWs are known to become more compressible at sub ion scales, although it is unclear whether they can fully account for the observed compressibility. Other modes such as the ion Bernstein wave or the KSW could potentially contribute to the compressibility (Lacombe et al. 2017 \emph{accepted to APJ}) if they exist in the plasma and can survive without being damped. Predictions based on a pure critically balanced cascade of KAWs or whistlers yield the same value for the spectral index of -7/3 \citep{Schekochihin2009,Chen2010} shallower than measured here. This has been attributed to the influence of Landau damping \citep{Howes2008} or intermittency \citep{Boldyrev2012a} where these predictions give a spectral index of -8/3 close to that measured here.

While perpendicular propagating fluctuations such as the KAW, ion Bernstein waves, or even kinetic slow waves may be incoherent in the two components perpendicular to the magnetic field, they would be expected to be coherent in the two components of the magnetic field which are perpendicular to their wavevector $\mathbf{k}$. However, for this interval $\mathbf{k}$ cannot be determined using multi-spacecraft observations since the {\it{Cluster}} separations are much larger than the scales of interest. One possibility would be to use Taylor's hypothesis, however at these scales whistler waves are highly dispersive and Taylor's hypothesis breaks down \citep{Howes2014a}. Moreover, even for KAWs which are not as dispersive as whistlers the wavevector can still propagate at moderate angles ($\sim 30^{\circ}$) with respect to the bulk flow \citep{Sahraoui2010a,Roberts2013} thus even when Taylor's hypothesis is valid it can have significant uncertainties \citep{Narita2017}. A final possibility would be to use the Poynting vector as a proxy for the wavevector $\mathbf{S}=\mathbf{E}\times\mathbf{B}/\mu_{0}$ but this method would require accurate measurements of the electric field.

 In this study we have examined the magnetic field power spectra at electron scales using the data from {\it{Cluster}} STAFF-SCM when in burst mode. However it is clear that a more sensitive instrument is required especially down below the Taylor shifted electron scales and past the Debye length scales. The data we have analyzed is from a very special event where burst mode data were available and where the turbulence has sufficiently high amplitude for the fluctuations to be measured at electron scales. It seems that the event includes an interaction between two streams of plasma, which may explain the high amplitudes of the turbulence. During this interval we have used wavelet coherence techniques to analyze the high frequency magnetic fluctuations and found several different fluctuation types. At electron scales parallel whistler waves can be seen and coherent structures can also be seen down to the electron characteristic scales. The morphology of the spectrum at electron scales is influenced strongly by the presence of sporadic and intermittent whistler waves, as well as coherent structures.  

Different spectral shapes are seen to be related to the coherence properties of the time series, with spectral knees and steep breaks being related to sporadic whistler waves. Meanwhile the underlying incoherent fluctuations show a similar universality as in \cite{Alexandrova2009,Alexandrova2012}, where no parallel whistlers were present in the intervals surveyed. Therefore the spectra of \cite{Alexandrova2009,Alexandrova2012} contain only incoherent fluctuations and coherent structures. The whistler waves observed in this study are short-lived of the order of seconds but can have a large effect on the shape of the power spectrum. We have shown four examples of the magnetic field power spectrum where the presence of whistler waves for $20\%$ of the 30 second spectra give a large spectral knee in the global spectra. When the abundance of whistler waves is smaller ($\sim 10\%$) the global spectra shows a clear break and a steepening. Meanwhile, the background spectra of incoherent fluctuations appear to have a curved exponential cutoff. Additionally, coherent structures contribute power over a large range of scales and the presence of only $\sim5\%$ strengthen the exponential shape of the spectrum.

Future work will involve looking at many more data sets with a variety of plasma parameters (e.g. $\beta$, solar wind velocity) to see the effects on the resulting spectra and the coherence. Moreover these studies should be performed simultaneously at fluid, proton and electron scales to understand the entire spectrum of plasma turbulence. This work also highlights the need for extremely sensitive electromagnetic field data with high time resolution which is far from Earth's foreshock. The new proposed European Space Agency M4 mission; THOR (Turbulence Heating ObserveR \cite{Vaivads2016}) plans to be able to obtain such detailed measurements.

\section{Acknoledgements}
All {\it{Cluster}} data are obtained from the ESA {\it{Cluster}} Science Archive: http://www.cosmos.esa.int/web/csa. We thank the FGM, CIS, PEACE, and STAFF instrument teams and the CSA team. We acknowledge the Coordinated Data Analysis Web (CDAWeb) service (https://cdaweb.sci.gsfc.nasa.gov/) for providing Wind data. OR and DP are funded by an ESA Science Research Fellowship, and acknowledge several helpful discussions of the joint ESTEC/ESAC heliophysics group. PK’s work is supported by the PAPIIT grant IA104416. OA thanks CNES for financial support for research on {\it{Cluster}}. The work of LT is supported by a Marie Sk\l{}odowska-Curie Individual Fellowship (grant agreement No 704681). OR and OA acknowledge illuminating discussions with C. Lacombe and L. Matteini.

\bibliographystyle{plainnat}
\bibliography{library}

%\end{thebibliography}



\end{document}